\documentclass{aa}  

\usepackage{graphicx}
\usepackage{txfonts}
\usepackage{color}
\definecolor{gray}{rgb}{0.5,0.6,0.7}
\definecolor{raspberry}{rgb}{0.7,0.2,0.4}
\definecolor{emerald}{rgb}{0,0.61,0.47}

\newcommand{\hi}{H\textsc{i}}
\begin{document}

   \title{Dark matter fraction derived from the M31 rotation curve}

   \titlerunning{M31 dark matter content}

   \author{F. Hammer\inst{1}, Y. B. Yang\inst{1}, P. Amram\inst{2}, L. Chemin\inst{3},  G. A. Mamon\inst{4}, J. L. Wang\inst{1}, I. Akib\inst{1}, Y. J. Jiao\inst{1}, \and H. F. Wang\inst{6} 
          }
\authorrunning{F. Hammer et al.}
   \institute{LIRA, Observatoire de Paris, PSL University, CNRS, Place Jules Janssen, Meudon, France
              \email{francois.hammer@obspm.fr}
          \and Aix-Marseille Univ., CNRS, CNES, LAM, 38 rue Fr\'ed\'ric Joliot Curie, 13338 Marseille, France       \and Instituto de Astrofisica, Facultad de Ciencias Exactas, Universidad Andres Bello, Fernandez Concha 700, Las Condes, Santiago RM, Chile
            \and  Institut d'Astrophysique de Paris (UMR7095: CNRS \& Sorbonne Universit\'e), 98 bis Bd Arago, 75014, Paris, France
          \and Dipartimento di Fisica e Astronomia "Galileo Galilei", Universita di Padova, Vicolo Osservatorio 3, I-35122, Padova, Italy
             }

   \date{Received October 25, 2024; Acepted November 30, 2024}

 
  \abstract
   {Mass estimates of a spiral galaxy derived from its rotation curve must account for the galaxy's past accretion history. There are several lines of evidence indicating that M31 experienced a major merger 2 to 3 Gyr ago. In this work, we generated a dynamical model of M31 as a merger remnant that reproduces most of its properties, including from the central bar to the outskirts. The model accounts for M31's past major merger and reproduces the details of its rotation curve, including its 14 kpc bump and the observed increase of velocity beyond 25 kpc. We find non-equilibrium and oscillatory motions in the gas of the merger-remnant outskirts caused by material in a tidal tail returning to the merger remnant. A total dynamical M31 mass of 4.5 $\times 10^{11} M_{\odot}$ within 137 kpc was obtained after scaling it to the observed HI rotation curve. Within this radial distance, we find that 68\% of the total dynamical mass is dark.}


   \maketitle
%

\section{Introduction}
The Andromeda galaxy (M31) is our nearest giant spiral neighbor. It has been considered as a twin of our Galaxy for decades, and as such, it offers a precious external view to our home. M31 has played a considerable role in establishing the modern cosmology, including in determining the difference between galaxies and Galactic nebulae as well as the presence of dark matter (DM) in the galaxy halo. The latter discovery took advantage of the very large sky angle displayed by the M31 disk, which is seen almost edge-on, allowing for easy measurement of its rotation from radial velocities. \citet{Lundmark1925} was one of the first to measure the rotation curve (RC) of M31, followed by \citet{Babcock1939}, and then \citet{Mayall1951}. They had already observed that it remained flat up to almost 20 kpc. Later, \citet{Rubin1978} and \citet{Bosma1978} showed that several spiral galaxies had flat RCs, including up to 35 kpc for M31.  

Considerable theoretical and observational progress has been accomplished since the late 1970s. Formerly, it was believed that disk galaxies were formed by an initial collapse \citep{Eggen1962}. Because thin disks cannot support significant merger events without being destroyed \citep{Toth1992}, the common consensus was that giant galactic disks have a stable rotation since very early epochs. This conjecture faced the problem of the angular momentum necessary to support disk galaxies \citep{Navarro2000}, that is, giant disks generated from cosmological fluctuations were too small. However, accurate comparisons between distant and nearby galaxies \citep{Hammer2009,Delgado2010} indicated that about 50\% of spiral galaxy progenitors were in a merger phase 6 Gyr ago. This can be understood because gas-rich mergers are very efficient at reforming a disk in the remnant \citep{Hopkins2009}, which also solves the disk angular momentum acquisition problem. This especially concerns M31, for which the accumulation of observations have revealed the giant Stream in its halo \citep{Ibata2001}, the past 2-3 Gyr star formation events \citep{Williams2017}, the stationary 10 kpc ring \citep{Dalcanton2012,Lewis2015}, and the unexpectedly large velocity dispersion of stars older than 2 Gyr \citep{Dorman2015}. All of these exceptional features have been reproduced by a gas-rich 4:1 merger that has occurred 2 to 3 Gyr ago \citep[hereafter H18]{Hammer2018}, which confirms the conjecture that the Milky Way and M31 are not twin galaxies because they have very different histories and dynamical properties \citep{Hammer2007}. The robustness of the M31 modeling has been confirmed by its ability to reproduce different observational features discovered later \citep{Bhattacharya2021,Bhattacharya2023,Escala2022,Tsakonas2024}. 

Having a recent merger in M31 implies a disk in which stars and gas have only recently reached their present orbits, which may warrant disk equilibrium in the inner regions but possibly not in the outskirts. The goal of this paper is to identify the DM content of M31 that is expected from the M31 RC in the context of a recent major merger. Section~\ref{sec:M31_RC} compares the different M31 RCs extracted from the HI gas motions and shows that they can be reconciled together. Section~\ref{sec:merger_RC} describes the RCs derived from M31 major merger models, how they can account for asymmetric disk effects, and which requirements on the DM content are necessary to best reproduce the M31 RC.  Section~\ref{sec:equilibrium} describes at which conditions disk stars and gas can be considered at equilibrium, which is based on orbit eccentricities, on the relative amplitude of radial and vertical velocities, and on observed differences between approaching and receding sides of the RC. Section~\ref{sec:discussion} compares our results with the literature studies on M31's mass, and then with the recently determined Milky Way RC, it also shows the limits of this study.
   
   \begin{figure}
   \centering
\includegraphics[width=8.5cm]{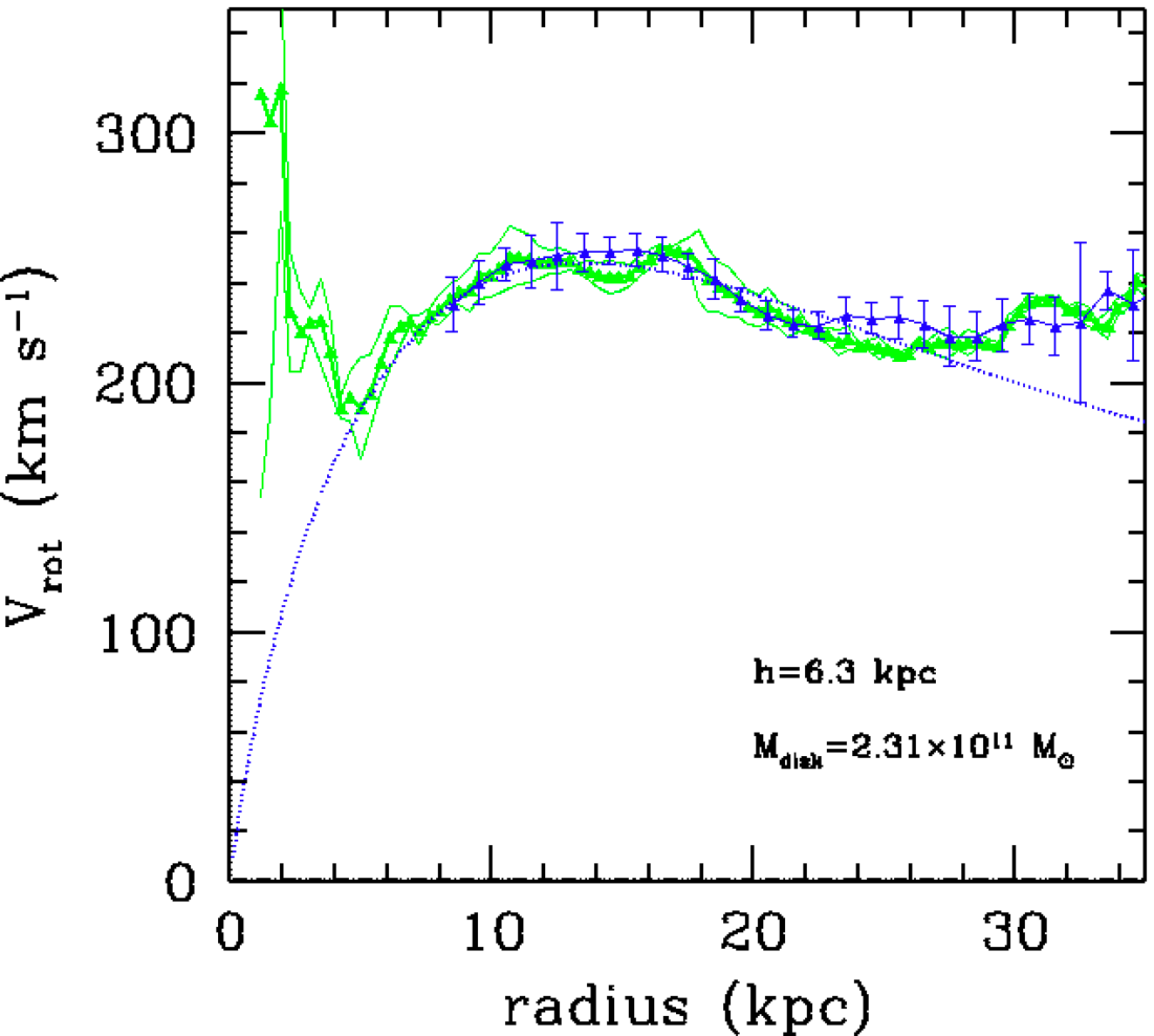}
      \caption{Comparison between \citet{Chemin2009} (green lines) and \citet{Corbelli2010} (blue dots) RCs. The former has been corrected by a factor of 0.94 to match the latter as best as possible. The dotted blue line shows the expected RC for a pure axisymmetric disk with a scale length of 6.3 kpc and a mass of 2.31 $\times 10^{11} M_{\odot}$.
              }
         \label{RC_Chemin_Corbelli}
   \end{figure}
%

\section{Rotation curve of M31}
\label{sec:M31_RC}
The best HI RCs of M31 have been derived by \citet{Chemin2009} and \citet{Corbelli2010}. The former study is based on deep 21cm observations made at the Synthesis Telescope and the 26 m antenna at the Dominion Radio Astrophysical Observatory, and the latter is based on a deep wide-field H I Westerbork and Green Bank Telescope survey of M31. Both surveys of the atomic HI disk of M31 have been made with a uniformly high sensitivity and resolution. Not only were the data taken at different radio telescopes, but their analysis methods also differ. The former study used terminal  
velocities derived from a multiple Gaussian peak analysis to get the main velocity field, 
while the second considered intensity-weighted mean velocities. 
Both studies applied a tilted-ring model to their respective \hi\ velocity field to infer the RC and warp parameters of the disk. Figure~\ref{RC_Chemin_Corbelli} compares the two RCs that show many similarities from 8 to 35 kpc. 
By construction, the curve of \citet{Chemin2009} exceeds that of \citet{Corbelli2010}, but both results remain very consistent within the quoted uncertainties, indicating that the terminal and average velocities trace the same gas component through most of the disk. 
Differences between the curves also come from small differences in the best fit parameters of the disk warping between the two studies.
We find that the best match between the two curves occurs when a correction factor of 0.94 is applied to the \citet{Chemin2009} RC, confirming that terminal velocities exceed intensity-weighted mean velocities.

\citet{Chemin2009} and \citet{Corbelli2010} RCs show a very similar behavior, with a sharp increase from $\sim$ 8 to 14 kpc, then a peak and a decrease beyond 16 kpc, which resembles a single bump. The major difference between the RCs occurs from $\sim$ 23 to 27 kpc, while beyond $\sim$ 28 kpc, both RCs show a further increase in velocity. Figure~\ref{RC_Chemin_Corbelli} reveals that the 14-15 kpc bump can be reproduced by the effect of a single axisymmetric disk, which corresponds to a scale length equal to the bump peak divided by 2.2, that is, from 6.2 to 6.7 kpc. The latter value is in excellent agreement with M31 estimates from photometry \citep[and references therein]{Hammer2007,BlanaDiaz2017}, and it allowed us to suspect that the origin of the bump indicates a quite prominent disk. We note, however,  that the value adopted in Figure~\ref{RC_Chemin_Corbelli} is not realistic since having all the mass concentrated into a massive disk would generate an extremely unstable disk \citep{Ostriker1973}. 

\begin{figure*}
   \centering
\includegraphics[width=17cm]{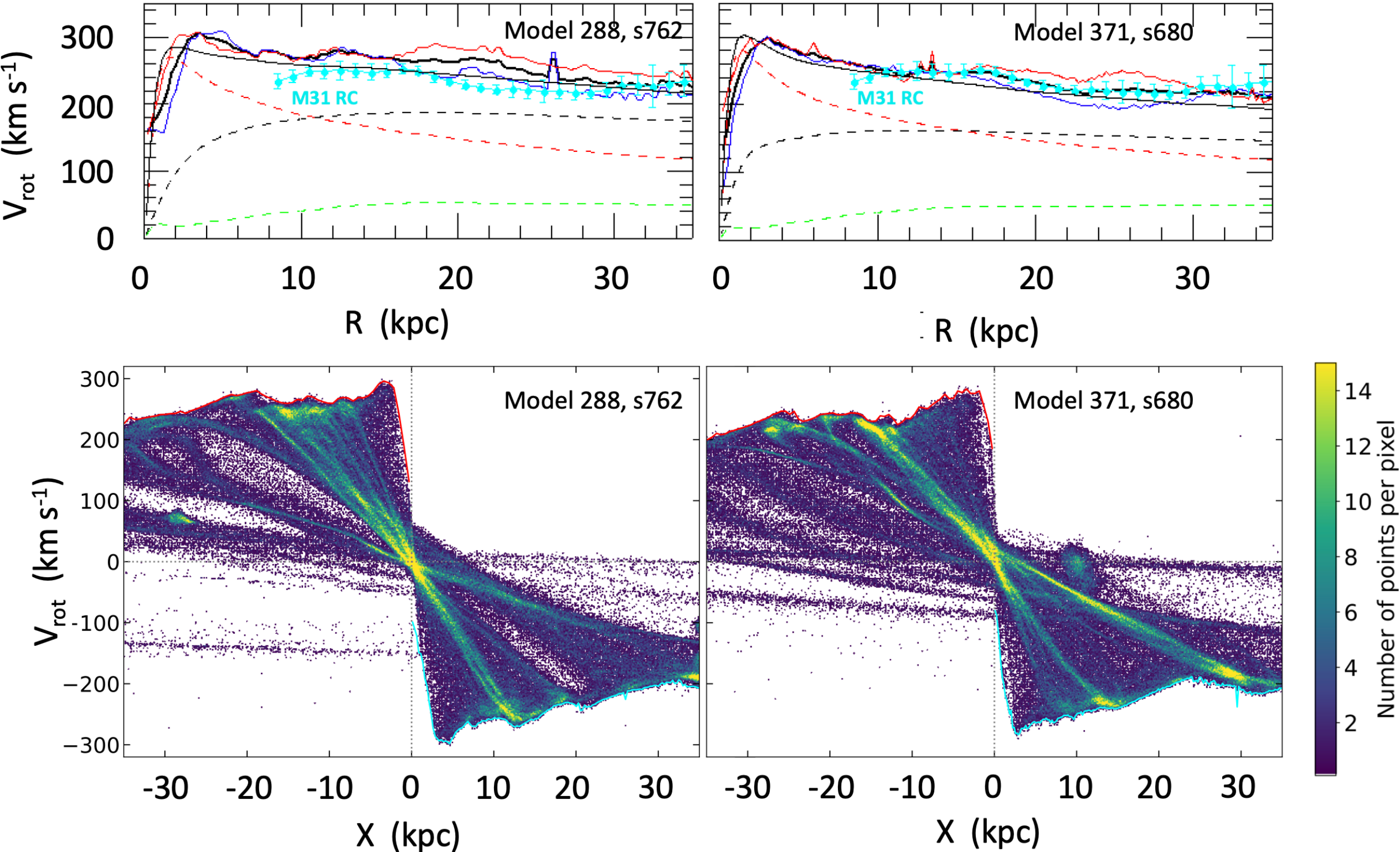}
      \caption{Derivation of the terminal velocity for two M31 models based on the maximal velocity method of the  position-velocity diagram \citep{Chemin2009} and comparison with the observed M31 RC (see Sect.~\ref{sec:M31_RC}).
Top panels:  RC derived from the maximal velocity method for models 288 (left) and 371 (right). The blue, red, and black lines represent RCs of the approaching and receding sides and the average. The observed M31 RC is shown by cyan dots with error bars and a cyan line. For illustration, the contribution of stars, DM, and gas to the modeled RCs are given by red, black, and green dashed lines, assuming a spherical symmetry, and the smooth black line corresponds to their summation. Bottom panels:  Position-velocity diagram plotted as a kernel density plot for gas particles with T $\le$ 20000K  (green dots) along the receding (X$<$ 0) and approaching (X $>$0) sides, with red and blue lines representing the maximal velocities that are also shown in the top panels. The figures also account for the gas particle densities in the phase diagram, which could be reasonably well compared to Figure 5 of \citet{Chemin2009}. In all panels, there are a few spikes that are due to single particles, which have been removed for further analyses (see Figure~\ref{RC_asymmetric}). 
}
         \label{Phasediagram_Vmax}
   \end{figure*}

Because the two M31 RCs from \citet{Chemin2009} and \citet{Corbelli2010} are so similar after re-scaling, we chose to generate a M31 RC that is the average of the two after correcting the first one by a factor of 0.94. We first interpolated the \citet{Chemin2009} RC to the same intervals as that of \citet{Corbelli2010}, reevaluating velocities and errors directly from the data. To estimate the final uncertainties, we quadratically added errors from \citet{Corbelli2010} and the systematic uncertainties coming from the difference between the two studies after applying the correction factor of 0.94 to the \citet{Chemin2009} RC.
In the following, we compare this averaged M31 RC (see cyan line and error bars in Figures~\ref{Phasediagram_Vmax} and ~\ref{RC_asymmetric}) to the RC derived from simulated galaxies made by the H18's hydrodynamical simulations.

%

\section{Rotation curves from merger modeling}
\label{sec:merger_RC}
H18 presented a set of five simulated models of M31 that provided very similar RCs (see their Appendix A), though they did not compare them to observational data. Since the stellar content of M31 is relatively well known ($M_{stars}$= 1.1 $\times 10^{11} M_{\odot}$), we varied the DM content to best match the observed RC. The top-left panel of Figure~\ref{Phasediagram_Vmax} indicates that for H18's model 288\footnote{We first chose model 288 because it provides a HI disk with a size that is comparable to the one observed (see H18's Figure 1).}, the rotational velocity is greater than the one observed (compare the full black line to the cyan points). We compared the rotational velocity of the model 288 with the observed velocity at 20 kpc. The choice of 20 kpc was guided by the fact that (1) there, the RC is not affected by the axisymmetric disk component (2) nor by non-equilibrium effects occurring at R > 25 kpc (see Section~\ref{sec:equilibrium}) and (3) that at 20 kpc the DM component fully dominates the baryonic component (compare the dashed black and red lines). This choice leads to a mass within 20 kpc that is close to the quadratic sum of the square velocities of the DM and baryonic components. This provides, quite accurately, a factor of 1.6 to be applied on the DM content of model 288, thus warranting that at 20 kpc the model fits the observed RC.
The latter factor led us to generate a new model, namely model 371\footnote{The properties of the new model are presented at the end of this section.}. The top-right panel of Figure~\ref{Phasediagram_Vmax} shows that the observed rotational velocity is quite well reproduced by the new model 371 (compare the full black line to the cyan points).

The bottom panels of Figure~\ref{Phasediagram_Vmax} show the position-velocity diagram provided by the HI gas particles (T$\le$ 20000K) for model 288 of H18 and for the new model 371, for which we decreased the DM mass by a factor of 1.6. The red and blue curves in the top panels of Figure~\ref{Phasediagram_Vmax} give the extremal velocities derived from the phase diagram (see bottom panels of Figure~\ref{Phasediagram_Vmax}) after correcting them for a constant inclination (i= 77 degrees). 

\begin{figure*}
   \centering
 \includegraphics[width=17cm]{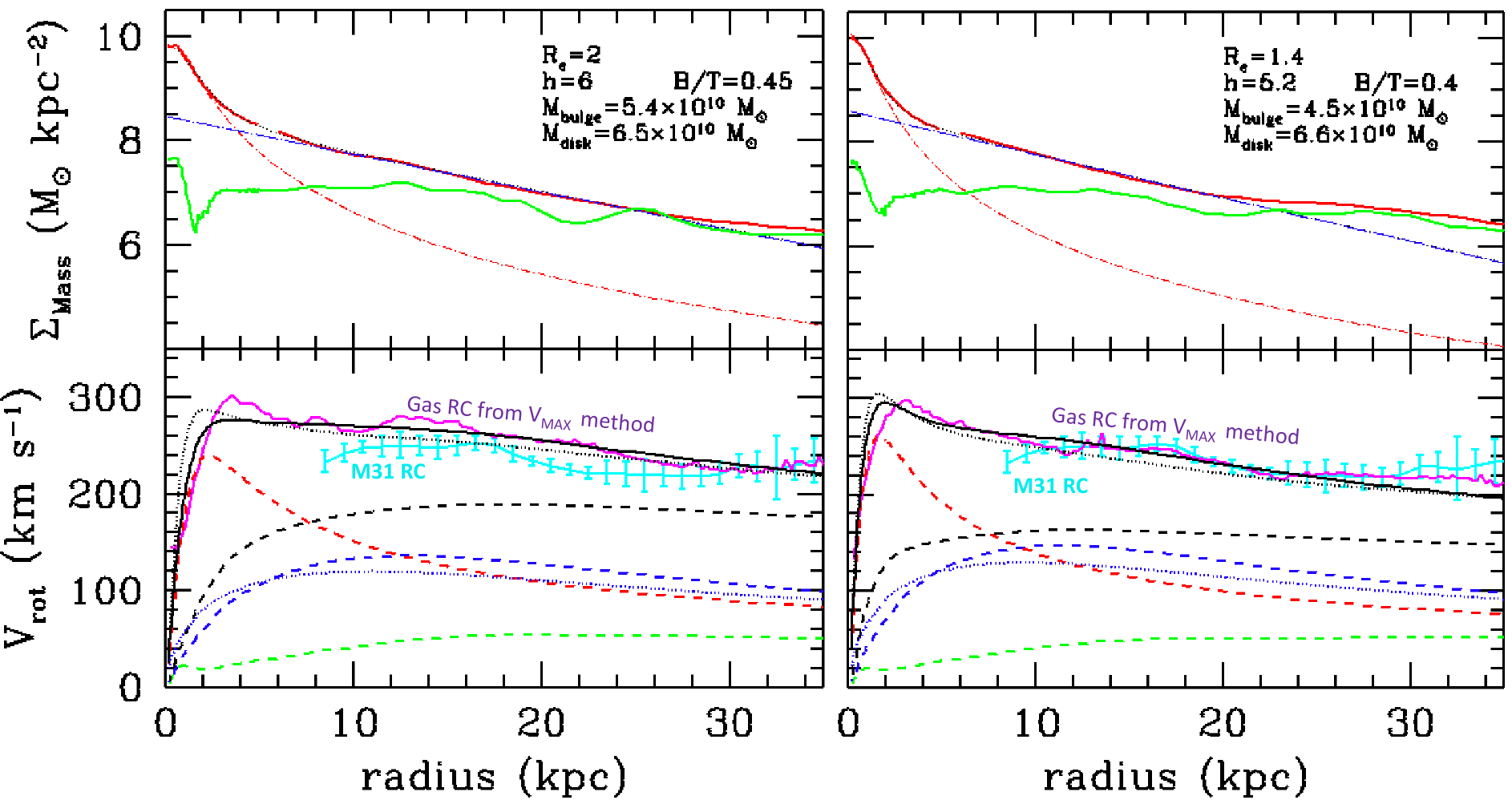}
  \caption{Derivation of the M31 modeled RC after decomposition of the baryonic component into a disk and bulge component for models 288 (left) and 371 (right). Top panels: Stellar mass surface density in logarithmic scale (red solid line) that is decomposed into a bulge component (red dashed line) and a disk (blue long-dashed line) together with that of the HI gas (green dashed line). The bulge and disk mass distributions result from a Plummer and exponential disk profile, respectively, and the characteristic numbers are given in the top right of each panel. Bottom panels: Same as the top panels of Figure~\ref{Phasediagram_Vmax} except that the black line represent the RC obtained after accounting for the bulge (red dashed line), axisymmetric disk (blue dashed line), and DM (black dashed line) components, respectively. The RC described by the maximal velocity method from Figure~\ref{Phasediagram_Vmax} is displayed with a magenta line , while the M31 RC is shown by a cyan line. The blue and black dotted lines illustrate the resulting RCs when the disk is (mis)-represented by a spherical component.
  } 
              \label{RC_asymmetric}
    \end{figure*}

Section~\ref{sec:M31_RC} (see also Figure~\ref{RC_Chemin_Corbelli}) indicates the presence of a large bump in the RC that is likely associated with the impact of the axisymmetric disk component of M31. To account for this, the top panels of Figure~\ref{RC_asymmetric} show the radial evolution of the stellar surface density profile (red solid line), which can be decomposed into a bulge (red dashed line) and a disk (blue long-dashed line), together with that of the HI gas (green dashed line). Although the stellar profile of the outskirts is underestimated, particularly for model 371 beyond 22 kpc\footnote{We verified that adding the corresponding missing baryonic mass does not change the RC beyond 22 kpc by more than 2 km$s^{-1}$ at 35 kpc and can be neglected. We also found negligible changes when passing from a thin to a thick disk for calculating the circular velocity.}, we may use this decomposition to predict the M31 model RCs in order to explicitly account for the asymmetric disk component. The bottom-left panel of Figure~\ref{RC_asymmetric} confirms that model 288 overestimates the observed velocity amplitude of the HI M31 RC (reduced $\chi^2$= 9.3), while model 371 provides a much better match between the HI observations and gas modeling (reduced $\chi^2$= 2.2, see bottom-right panel of Figure~\ref{RC_asymmetric}). We further note that the observed increase of the gas rotational velocity beyond 28 kpc (see Section~\ref{sec:M31_RC}) is well reproduced by model 371 but not by model 288.

%
   %

Another important characteristic of the HI M31 RC is the discrepancies in velocity amplitude between approaching and receding RCs at different radii. This indicates that there are some perturbations of the disk dynamics. Such discrepancies were observed by \citet{Chemin2009} between 8 and 12 kpc (inner discrepancy) and then between 17 and 21 kpc (outer discrepancy). We first noticed that while the \citet{Corbelli2010} RC shows a similar behavior, the discrepancy between both sides has a smaller amplitude (15 versus 30 km$s^{-1}$) and slightly different radii: from 11 to 13 kpc and from 22 to 31 kpc for the inner and outer discrepancies, respectively. We suggest that the observational differences are related to the different methodologies used by both studies to recover the RCs (see Section~\ref{sec:M31_RC}). \citet{BlanaDiaz2017} proposed that the 10 kpc ring could be related to the bar's outer Lindblad resonance, for which the bar pattern speed is in agreement with dynamical results from \citet{BlanaDiaz2018}. This suggests that the inner discrepancy that occurs in the same radial range could also be associated with the bar. However, we still have to interpret the observed outer discrepancy between the receding and approaching velocities, which is predicted by both models 288 and 371 to occur between 17 to 30 kpc (see red and blue curves in the top panels of Figure~\ref{Phasediagram_Vmax}). We note that the modeled discrepancy in velocity reaches approximately a maximal amplitude of 40 km$s^{-1}$, which is quite similar to what has been found by \citet{Chemin2009} but significantly larger than results from \citet{Corbelli2010}. This could be due to the fact that we used the same methodology to extract modeled RCs as \citet{Chemin2009}.\\

\begin{figure}
   \centering
 \includegraphics[width=6.5cm]{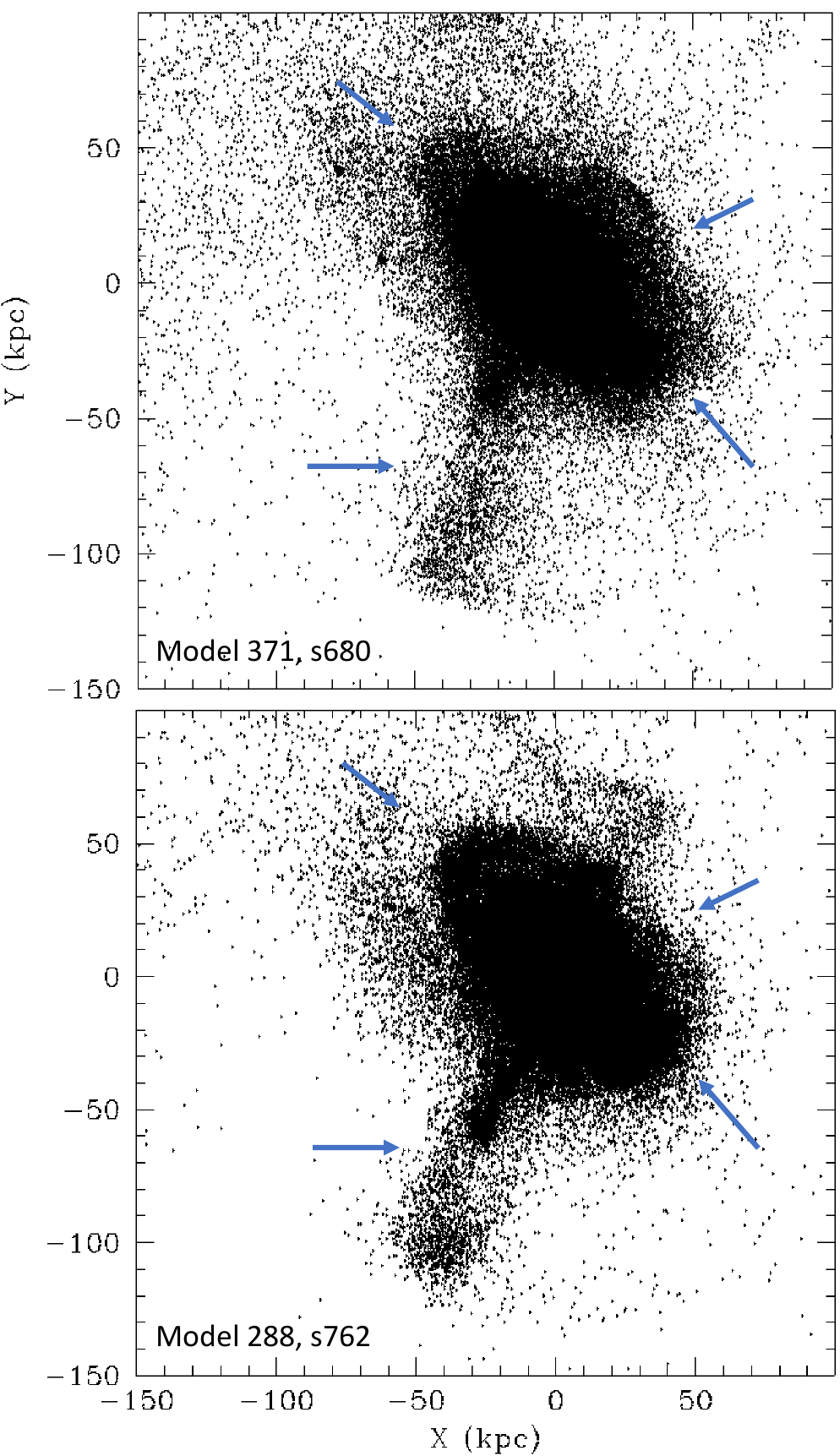}
   \caption{Sky projection of the stellar density of model 371 (top panel) that can be compared to model 288 (bottom panel) as well as other similar modeling made by H18 (see their Figure 8). The arrows indicate major features, from left to right and top to bottom: NE clump, W shell, G1 clump, and the giant stream. }
              \label{Mod371}
    \end{figure}

\begin{figure*}
   \centering
 \includegraphics[width=14cm]{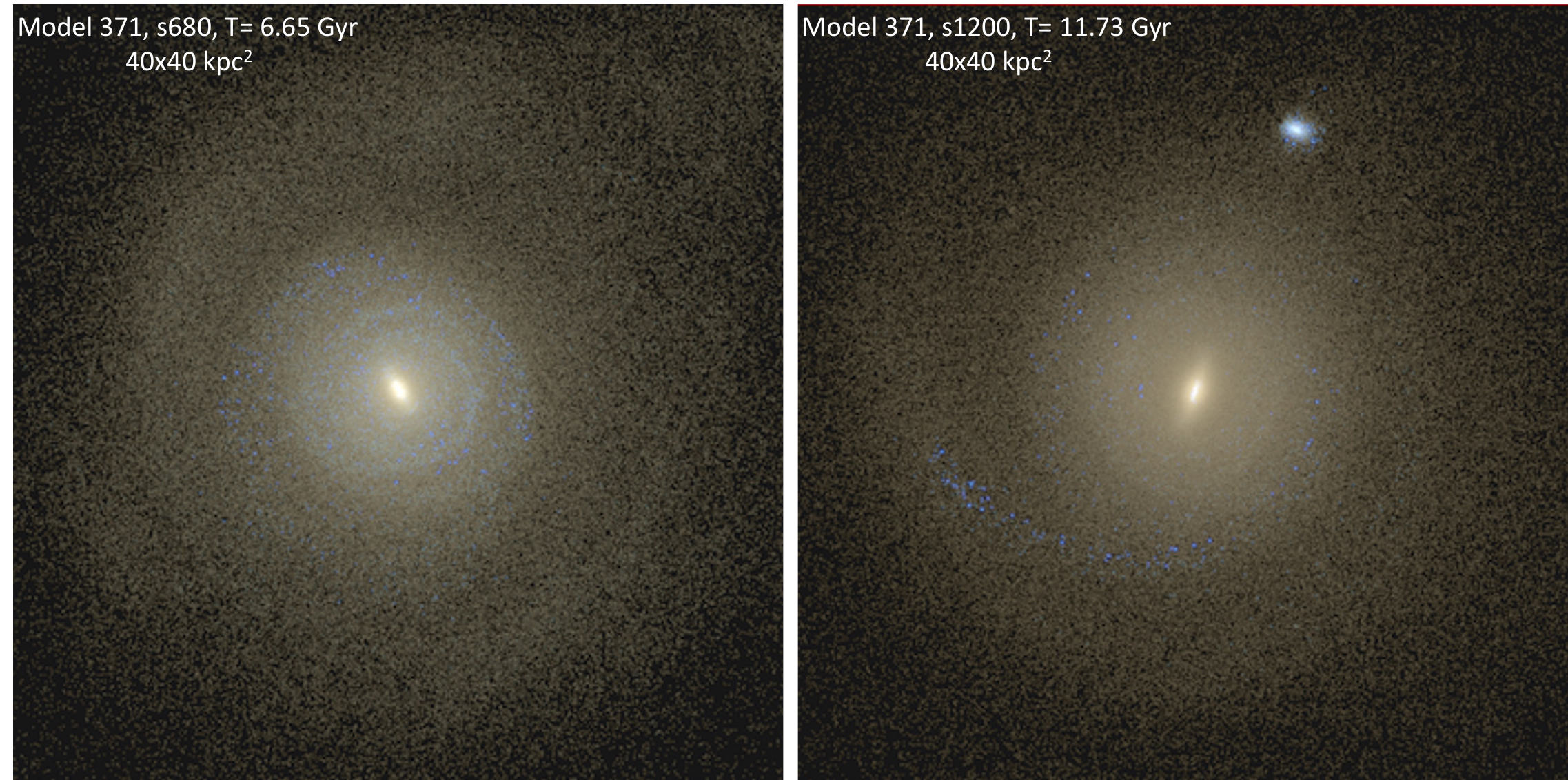}
   \caption{Stellar particle distribution of the face-on disk from model 371 within 40 kpc for snapshot 680 (left) and 1200 (right). The color of the image is
based on assuming each newborn stellar particle evolves as a simple stellar population and has been observed with the filters of u, g, and z of the Sloan Digital Sky Survey. For example, the blue colors in the images indicate a stellar age of about 100 Myr.}
              \label{fig:faceondisks}
    \end{figure*}

The above suggests that to best reproduce the observed M31 RC, the required missing mass has to be reduced from that considered by H18 (see also Section~\ref{sec:discussion}). However, before such a conclusion could be reached, we had to verify whether the main successes of H18 modeling were kept after reducing the DM content by a single factor of 1.6 (model 371). Figure~\ref{Mod371} shows the overall stellar distribution predicted by model 371, and it indeed reproduces the M31 disk outskirts, including the giant stream, shells, and clumps, in a manner similar to model 288 (see Figure 8 of H18). We also verified that the age-velocity dispersion relation established by \citet{Dorman2015} is also reproduced. The left panel of Figure~\ref{fig:faceondisks} shows the distribution of stellar particles in the face-on disk, and it illustrates that the new model 371 also reproduces the 10 kpc ring of star formation (see blue points that identify young stars) and the presence of a bar, which is slightly more prominent than in model 288. Furthermore, as done in H18, we verified that for a snapshot of model 371 taken 1 Gyr later (snapshot 780 instead of 680), both the outskirt morphological features and age-velocity dispersion relation are significantly diluted and merely reproduce the observations. The success of the new model 371 in reproducing M31's overall properties is not unexpected because these exceptional properties are caused by the interaction (including resonances) of the two disks of the merger interlopers,\footnote{We emphasize here that the bar occurs due to the resonance between the initial baryonic disks, while the giant stream is mostly made by particles coming from the secondary interloper.} that is, without much impact from the DM fraction.  \\

%
   %

%
   %


   \begin{figure}
   \centering
\includegraphics[width=8.5cm]{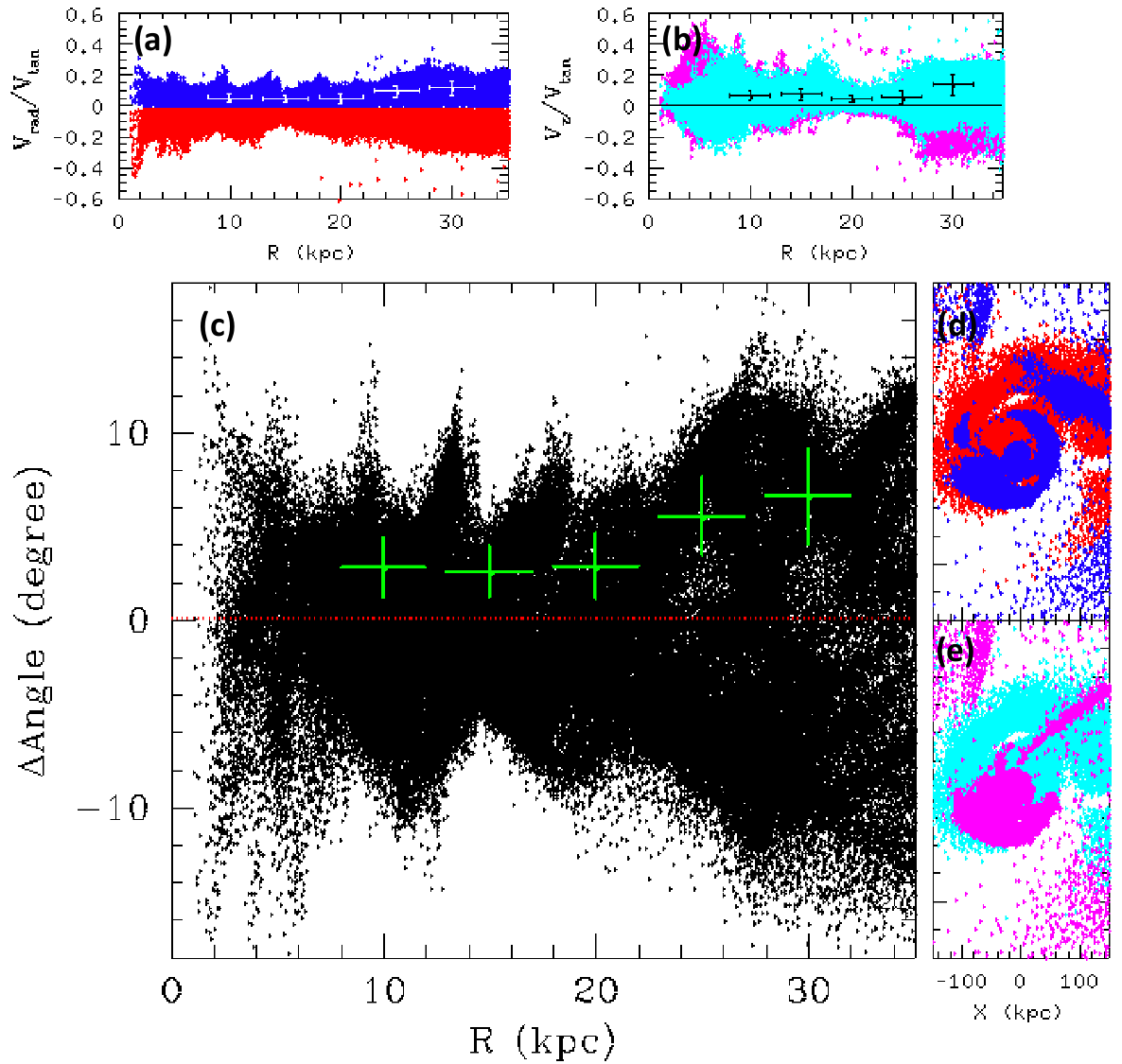}
      \caption{Oscillation behavior of gas particles in the merger model 371 and snapshot 680 (i.e., 2.6 Gyr after the merger). Panels a and b give the ratio of the radial (and azimuthal) velocities to tangential velocities,
respectively. In both panels, points and error bars give the median of the
absolute values at five different radii. The colors in panel b indicate whether
$\Delta\mathrm{Angle}$ is negative (magenta) or positive (cyan). Panel c shows the radial evolution of $\Delta\mathrm{Angle}$, which is the difference between the disk-projected velocity vector and that
of a pure circular motion. Points and error bars give the median of the
absolute values at five different radii. Panels d and e provide the gas
particle distribution within 300 $\times$ 600 $kpc^{2}$ and show the presence of a
tidal tail on the right sides whose angular momentum is slightly tilted versus
that of the remnant disk. The colors are the same as in panels a and b,
respectively.
              }
         \label{fig:Angle_s680}
   \end{figure}
%
   \begin{figure}
   \centering
\includegraphics[width=8.5cm]{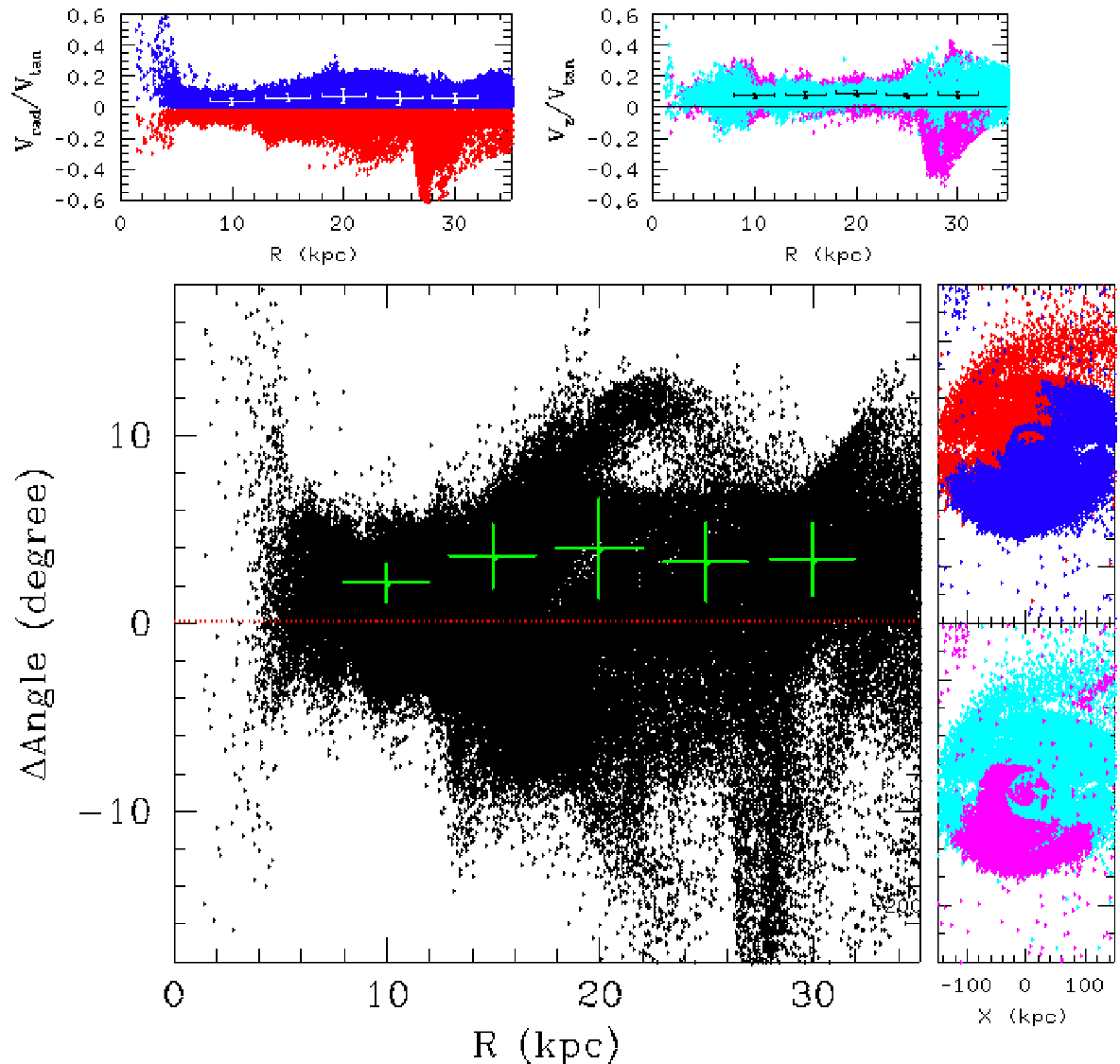}
      \caption{Same as Figure~\ref{fig:Angle_s680} but for snapshot 1200 (7.8 Gyr after the
merger) of model 371. The gas particles forming an almost vertical line with very negative values at R=27 kpc are caused by a tidal dwarf that is found superimposed to the disk (see right panel of Figure~\ref{fig:faceondisks}).
              }
         \label{fig:Angle_s1200}
   \end{figure}
%

\section{The radius to which the mass can be estimated in M31}
\label{sec:equilibrium}

Studies of RCs of remnant disks or disks that survive after a merger are not common in the literature. Here, we try to establish some simple figures to understand how such galactic disks can be stabilized over time. As for any dynamical system, the central parts can rapidly return to equilibrium. However, they can be affected by non-axisymmetric perturbations such as those caused by a bar, and this appears to be the case for M31 (see Section~\ref{sec:merger_RC}). The situation is different for the gas disk outskirts, which are particularly interesting here because they are the most sensitive to the presence of the DM that is expected to be dominant in the halo.

Non-axisymmetric perturbations of the outer M31 regions 
can be attributed to the impact of ancient tidal tails formed during the merger, in which part of the gas and stars may have reached their apocenter before returning to the remnant galaxy with relatively large velocities. Such a mechanism explains well the formation of the giant stream (see H18), which is made by particles forming loops around the M31 remnant after returning from a tidal tail associated with the secondary galaxy involved in the merger. The above result is a significant success of the M31 modeling because 
the latter structure is found to be composed by many substructures associated with the superposition of the loops that are seen edge-on \citep{Hammer2018,Escala2021,Tsakonas2024}. Each of the five models selected by H18 to reproduce M31 and its outskirts form two or even three tidal tails that are associated with the orbital path of the secondary galaxy. Each model also shows a tidal tail that is associated with the main galaxy and possesses a slightly tilted angular momentum compared to that of the remnant disk. The two above properties are shared by model 371 (see Figure~\ref{Mod371} and panels d and e of Figure~\ref{fig:Angle_s680}). 

Figure~\ref{fig:Angle_s680} provides a description of the mechanisms at work for the merger model 371. Panel (c) shows how the gas orbital 
motions differ from circularity, indicating an offset angle ($\Delta\mathrm{Angle}$) that increases with radii and reaches a median value close to 10 degrees at
30 kpc. Similar radial increases were found for the ratios between 
radial (panel a) and azimuthal (panel b) to tangential velocities, 
respectively. Panels (d) and (e) show a large scale map of the 
remnant gas that is dominated by a gigantic tidal tail (see right 
sides of each panel), which is directly linked to the disk. Figure~\ref{fig:Angle_s680} 
suggests that the radial increase of the non-circular motions are 
caused by particles returning from the tail. Moreover, one 
(the other) side of the disk has negative (positive) radial velocities, which supports well the fact that the disk is permanently shocked by gas particles infalling from the tidal tail. 
We note that it also explains the oscillatory behavior of the gas particles in
panel (c) as well as the differences between the approaching and receding RCs occurring mostly beyond 17 kpc (see the differences between the green lines in Figure~\ref{RC_Chemin_Corbelli} and between the red and blue lines in the top panels of Figure~\ref{RC_asymmetric}). 
To verify the above, it is useful to investigate the properties of model 371 but at the much later epochs for which it is expected to reach an equilibrium even at the disk outskirts. Figure~\ref{fig:Angle_s1200} is the same as Figure~\ref{fig:Angle_s680} but for the remnant taken at 7.8 instead of 2.6 Gyr after the merger. It indicates that most of the system appears relaxed. In particular, there is a flat distribution of the angle offset from circularity, the radial and azimuthal velocities, for which contribution becomes very small at all radii when compared to tangential velocities. 

   \begin{figure*}
   \centering
\includegraphics[height=10cm]{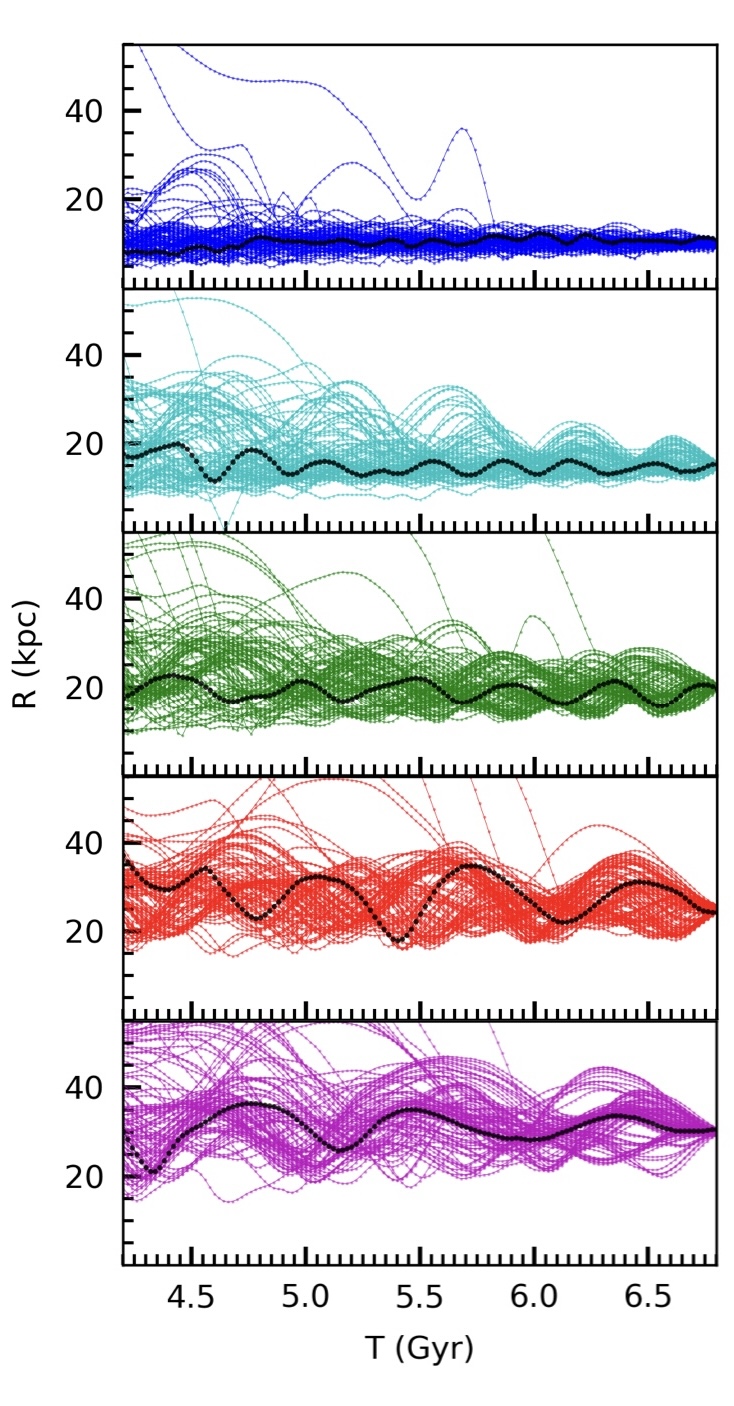}\hspace{1.cm}\includegraphics[height=10cm]{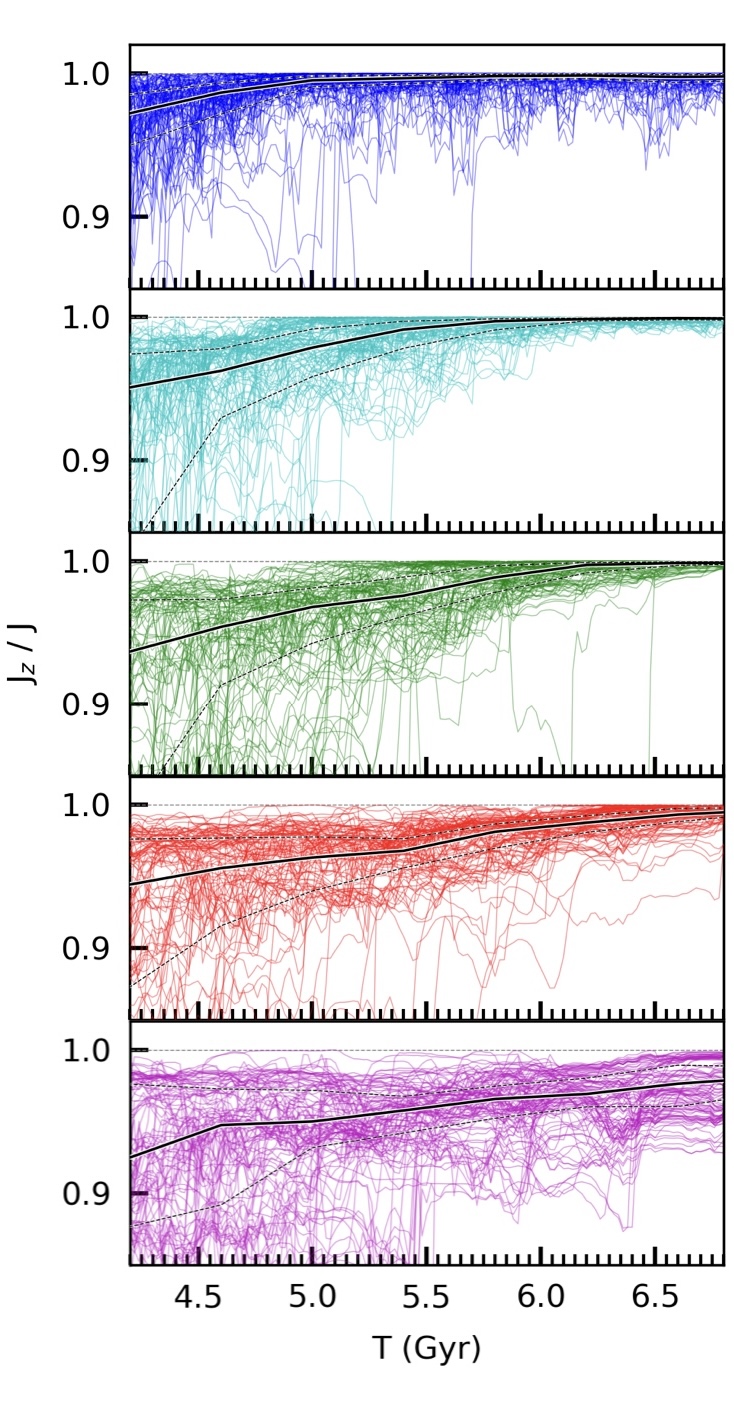}\hspace{1.cm}\includegraphics[height=10cm]{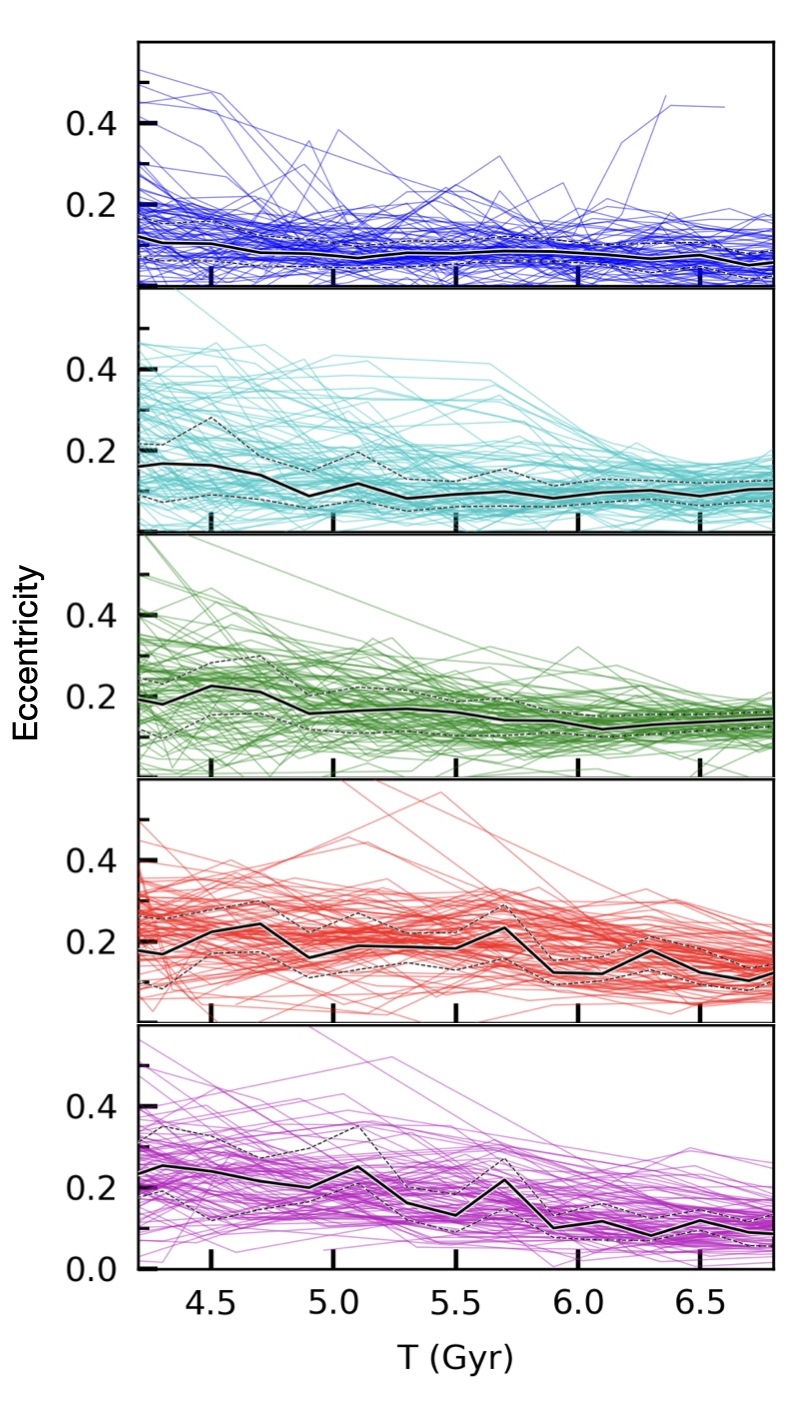}
   \caption{Temporal evolution from T=4.2 Gyr (merger epoch) to 6.8 Gyr
(present-day of the remnant M31) of randomly selected gas particle orbits
at five different areas of the disk. The areas correspond to 2 kpc
wide rings selected in the gas disk at T=6.8 Gyr (snapshot 680) with
radii of 10 (blue), 15 (cyan), 20 (green), 25 (pink), and 30 (magenta)
kpc, respectively. Left panel: Radial orbital evolution of the gas particles
from 10 kpc (top panel) to 30 kpc (bottom panel). The black dotted
lines illustrate the orbital paths of characteristic gas particles in each
panel. Middle panel: Ratio of the disk angular momentum ($J_{\rm z}/J$) to the
total angular momentum in the same rings as in the left panel. The
black solid lines give the median, and the black dashed lines give the 1$\sigma$ variations. Right panel: Same as the other panels but for the eccentricity evolution.
}
              \label{fig:time_evolution}
    \end{figure*}
%
   %

Figure~\ref{fig:time_evolution} describes the orbital evolution
from the merger (4.2 Gyr) to the present (6.8 Gyr) epoch. The leftmost and rightmost panels indicate that gas motions are increasingly eccentric, from the top to the bottom panels, for which gas particles are passing from pericenter to apocenter. At 10 kpc (top panel, blue) gas motions are mostly circular (same apocenter as pericenter), while for 
gas particles selected at 25 and 30 kpc (pink and magenta), the apocenter and pericenter may differ by more than 10 kpc. One may assume that the main perturbation occurred at the merger epoch and that the gaseous disk system gradually regains equilibrium after a certain number of orbital times \citep{Gnedin1999}. 
For example,  \citet{Gnedin1999} estimated that after a tidal shock, 
stars in a globular cluster retrieve the virial relation after three dynamical times, which we can translate as the number of orbital times for the disk. However, according to \citet{Gnedin1999}, a shocked system is still oscillating, and 10 to 15 dynamical times are necessary to reach full virial equilibrium. By similarity, this would mean that at 10 kpc, gas dynamics can be considered to fully be at virial equilibrium while still oscillating further away. 

However, the gas relaxation process behaves differently than expected based on stellar system interactions. After a major merger, the gas suffers from violent relaxation and a dissipative process, which leads to a faster relaxation process than that of a stellar system. It provides a short timescale for the gas-disk formation, and even in some cases, the gas disk is implemented just before the end of the merger process \citep{Barnes2002}. 
Figure~\ref{fig:time_evolution} illustrates how the gas returning from a tidal tail gradually forms a rotating disk.  \citet{Barnes2002} has shown that tidal tail gas particles initially side swipe 
the central disk and then follow eccentric orbits before settling down to the disk. The accreted gas intersects and shocks, which converts kinetic energy into heat that is radiated away through cooling \citep{Barnes1996}. Gas located in outer regions has more energy and less density, conditions in which the dissipation is less efficient than in the inner region. In other words, the gas reaches circular orbits preferentially in the inner region and gradually in the outer regions, as shown in Figures~\ref{fig:Angle_s680} and ~\ref{fig:time_evolution}.

In the central regions, the stellar bar plays an important role in draining the angular momentum to the gaseous disk; this 
has been extensively studied by \citet{Hopkins2009}. The
stellar bar has a different rotation compared to the gaseous disk, and this small offset exerts a torque on the gas that transfers its angular momentum into the system toward the central region. Since there is a bar in the M31, the process should also play a role in regulating the inside gaseous disk. In the outskirts, gas takes more time to adopt circular motions because it is permanently fed by the gas particles returning from the tail. Similar conclusions can be derived from the middle and right panels of Figure~\ref{fig:time_evolution}. The bottom panels indicate that gas particles have not fully reached orbital and circular motions inside the thin disk, for which $J_{\rm z}/J$ and eccentricity are equal to one and zero, respectively. The middle panels of Figure~\ref{fig:time_evolution} show that the gas has difficulty reaching precisely $J_{\rm z}/J$= 1 at the outskirts. This is because the cold gas has a sound speed of $\sim 16$km/s (for a gas temperature of 20000 K), which is much smaller than the rotation speed, so the gas pressure force is much smaller, and the gas will roughly follow the ballistic trajectory. Because of the above arguments, Figure~\ref{fig:time_evolution} confirms the conjecture of \citet{Gnedin1999} that virial equilibrium is not reached if the gas has orbited less than three (and five) times around the M31 center (i.e., beyond 25 kpc). This makes a calculation of mass from the virial theorem beyond that radius inapplicable. 

The top-middle panel of Figure~\ref{fig:time_evolution} also shows a very interesting property at 10 kpc because a few gas particles show almost periodic variations of their angular momentum during the whole 2.6 Gyr elapsed time since the merger. We suggest that this reveals the role of the
 bar, which has been proposed to be at the origin of gas shocks observed in M31's central regions \citep{Feng2024}.
 
Future works investigating the properties of the bar and how to model it would be very useful, as such investigations are beyond the scope of this paper. Preliminary results have shown that the modeled bar and its associated velocity field show strong variations with time, which could be associated with non-equilibrium conditions below 10 kpc. It is also possible that the central regions have an oscillatory behavior, suggesting a pseudo-equilibrium, that is, a system that is varying but retrieving similar conditions during periodic elapsed epochs.
 
   \begin{figure}
   \centering
 \includegraphics[width=8cm]{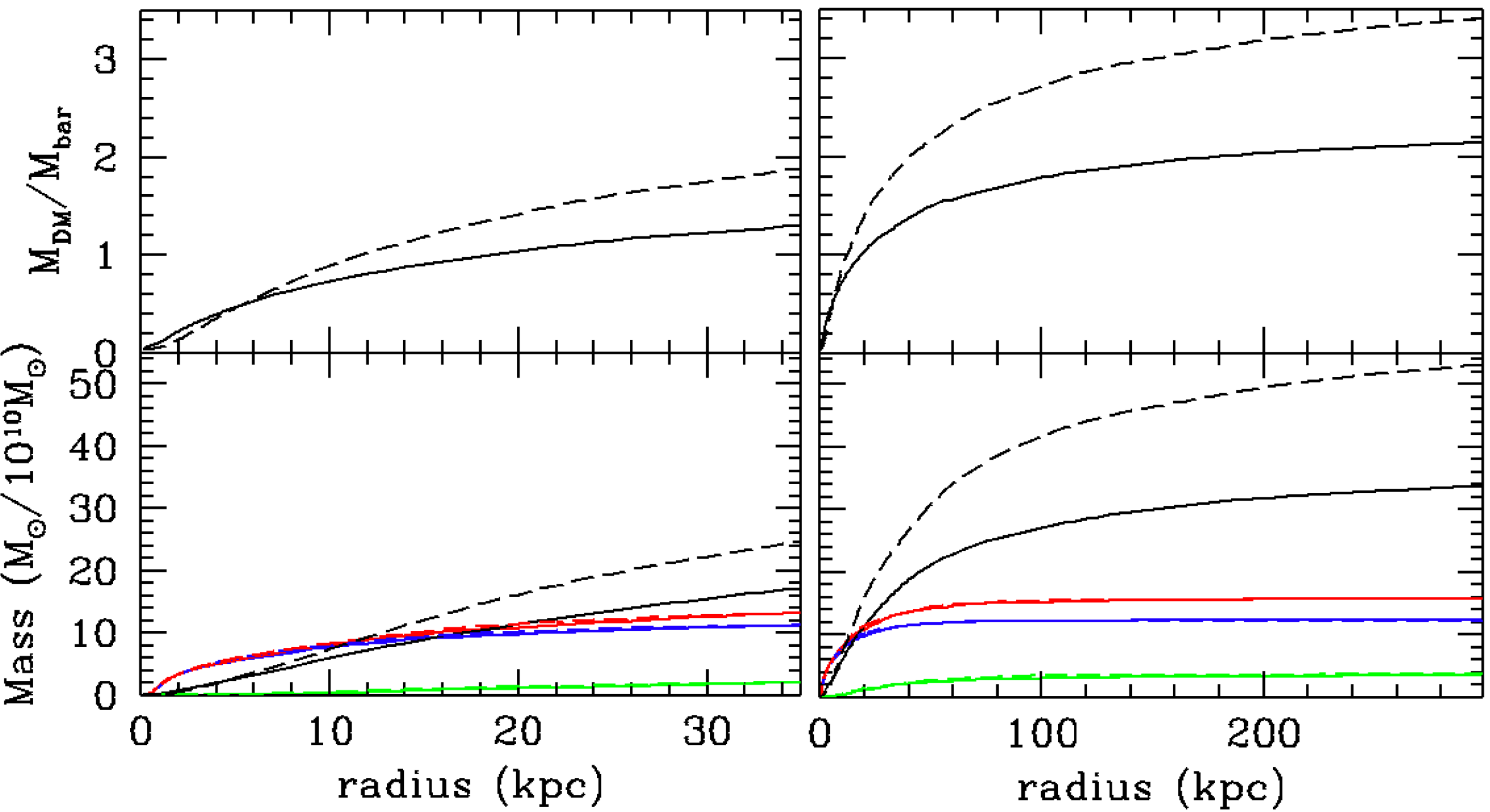}
   \caption{Radial distribution of DM and baryonic matter. The full lines show the radial DM distribution of model 371 (snapshot 680), within 35 kpc (left) and 300 kpc (right). Long-dashed lines
give the same for model 288 and snapshot 762. Top panels: Ratio of $M_{\rm DM}/M_{\rm bar}$ versus radius. Bottom panels: Evolution of DM (black lines),
baryon (red lines), stellar (blue lines), and gas (green lines) masses. }
              \label{fig:DM_distribution}
    \end{figure}

\section{Discussion}
\label{sec:discussion}
In this paper, we have shown that the M31 RC is reproduced by model 371, which is similar to M31 models developed by H18 but after reducing the DM mass by a factor 1.6. We have also shown that the former H18 models systematically overestimate the M31 RC because of their too large DM fractions (see left panels of Figures~\ref{Phasediagram_Vmax} and ~\ref{RC_asymmetric}). One may notice from Figure~\ref{RC_asymmetric} that the baryonic mass is also larger for model 288 than for model 371. We verified that model 288's RC still overestimates the observed RC after adopting the smaller baryonic mass associated with model 371.  

Since model 371 reproduces most of M31's properties as well as H18's modeling, we investigated how the M31 RC can predict the dynamical mass and to which limiting radius the virial relation can be retrieved. Both observations (see Figure~\ref{RC_Chemin_Corbelli}) and simulations (see left 
panels of Figures~\ref{Phasediagram_Vmax} and ~\ref{RC_asymmetric}) show a significant velocity increase beyond 25 kpc. However, Figure~\ref{fig:Angle_s680} indicates that at these radii, rotational velocities are affected by instabilities caused by gas particles returning to the merger remnant from 
a tidal tail. These particles also affect the orbital eccentricity as well as the radial and 
azimuthal velocities of the gas particles in the disk outskirts. The 
temporal evolution of gas particle orbits since the merger epoch (4.2 Gyr) confirms the above (see left panel of Figure~\ref{fig:time_evolution}). During the 2.6 Gyr elapsed time shown in Figure~\ref{fig:time_evolution}, gas particles at disk outskirts have experienced only a small number of orbits, two to four at 30 and 25 kpc, respectively. These numbers are too small to warrant virial equilibrium 
with the M31 remnant gravitational potential. 
We consider that model 371 represents well the matter 
distribution in M31. Figure~\ref{fig:DM_distribution} gives the radial distribution of the DM and of the baryonic components as well as the $M_{\rm DM}/M_{\rm bar}$ ratio. It indicates a baryonic component that represents 32\% of the M31 dynamical mass when considering it in a volume limited by 
$R_{\rm 200}$= 137 kpc, a DM, and a total mass of M31 of $M_{\rm 200}$= 2.95 $\times 10^{11} M_{\odot}$ and $M_{\rm tot}$= 4.5 $\times 10^{11} M_{\odot}$, respectively.\footnote{In this paper, $R_{\rm 200c}$ is the virial radius for which the enclosed DM mass, dubbed $M_{\rm 200c}$, corresponds to an overdensity of 200 times the critical density. The term $M_{\rm tot}$ accounts for DM and baryons, including 0.35$ \times 10^{11} M_{\odot}$ of the neutral gas (see Figure~\ref{fig:DM_distribution})}  

However, there is an important concern related to the baryonic mass, which may be different from one study to another. The bulge mass chosen for fitting the stellar mass distribution of model 371 is very close to that of the accurate modeling of M31's central region \citep{BlanaDiaz2018}: 4.5 versus 4.25 $\times 10^{10} M_{\odot}$. The total stellar mass of M31 in model 371 is 12 $\times 10^{10} M_{\odot}$, which is an intermediate value compared to \citet[9.5 $\times 10^{10} M_{\odot}$]{Chemin2009}, \citet[10.3 $\times 10^{10} M_{\odot}$]{Kafle2018}, and \citet[12.6 $\times 10^{10} M_{\odot}$]{Corbelli2010}. We verified that changing the baryonic content to values adopted in other studies would not change our conclusions about the DM total mass simply because the latter dominates the RC beyond 20 kpc (see bottom-left panel of Figure~\ref{fig:DM_distribution}). Given the above, we may consider that our mass determination accounts for all substructures of M31 and of its RC as well as for equilibrium conditions and that it supersedes the mass estimates made by \citet{Chemin2009} and \citet{Corbelli2010}.

\citet{Bhattacharya2023review} reviewed the different mass estimates for M31. There is quite a large variety of estimates because they are based on many different probes. However, very few of them \citep{Escala2022,Dey2023} have accounted for the fact that M31 has experienced a recent merger, which affects equilibrium conditions. For example, it might be difficult to consider orbits of probes lying at distances significantly larger than 30-40 kpc, as they would not have had time to experience more than one or two orbits since the merger time (see bottom-left panel of Figure~\ref{fig:time_evolution}). This especially concerns mass determinations based on dwarf galaxy satellites or distant globular clusters that predict total masses from 11.4 $\times 10^{11} M_{\odot}$ \citep[star clusters at distances larger than 50 and even 100 kpc]{Zhang2024}, 13.5 $\times 10^{11} M_{\odot}$ \citep[globular clusters up to 130 kpc]{Veljanoski2013}, and 14 $\times 10^{11} M_{\odot}$ \citep[dwarf galaxies within 300 kpc]{Watkins2010} to 19.9 $\times 10^{11} M_{\odot}$ \citep[all probes]{Sofue2015}. In the following, we restrict comparisons to previous works that accounted for kinematic probes at distances smaller than $\sim$ 35-40 kpc.

\citet{Dey2023} proposed an innovative method to estimate the mass of M31 based on the extent of the giant stream and associated shells (see also the pioneering studies of \citealt{Fardal2013} and \citealt{Escala2022}). They derived a mass for the DM halo of 6.3 $\times 10^{11} M_{\odot}$ within 125 kpc, which is twice our value (2.95 $\times 10^{11} M_{\odot}$ within 137 kpc). The study of \citet{Dey2023} is based on a progenitor of the giant stream with a much smaller mass ($10^{8} M_{\odot}$) than in the H18 modeling, and it is unclear whether this can affect their determined total M31 mass. Furthermore, \citet{Dey2023} acknowledged that their determined mass may overestimate the mass by a factor of two to three (see a detailed description in \citealt{Sanderson2013}), which would render their mass estimate consistent with that from model 371.

\begin{figure}
   \centering
 \includegraphics[width=8cm]{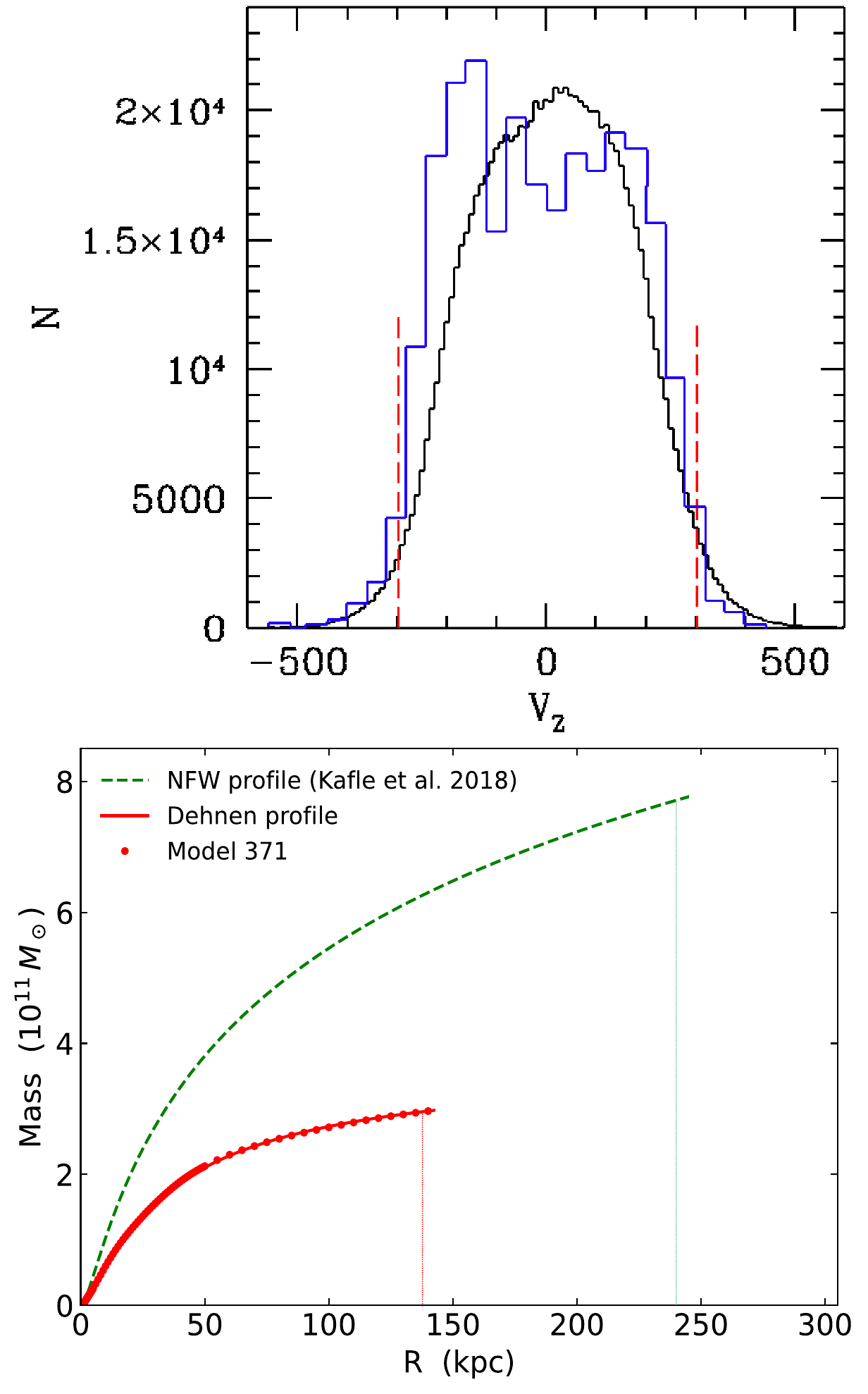}
   \caption{Comparison with \citet{Kafle2018}. Top panel: Distributions of PNe radial velocities (blue histogram) from \citet{Kafle2018} compared to predictions from model 371 (black histogram). The two red-dashed vertical lines show the limits chosen by \citet{Kafle2018} to calculate M31's escape velocity from PNe with absolute velocity values in excess of 300 km$s^{-1}$. Bottom panel: Cumulative mass $M_{\rm 200}$ for Model 371 (red points) fit by the function of \citet{Dehnen1993} (red line, see Eq.~\ref{eq:dehnen}) compared to that from \citet[gree-dashed line]{Kafle2018}. Vertical dotted lines indicate the corresponding $R_{\rm 200}$ radii: 137 and 240 kpc, respectively.
   }
              \label{fig:DM_density_Kafle}
    \end{figure}

\citet{Kafle2018} studied M31's planetary nebulas (PNes) in an attempt to derive the escape velocity profile of M31 and then its total mass. They found an escape velocity of 470 km $s^{-1}$ after assuming a distribution of stars with absolute values of radial velocities larger than 300 km $s^{-1}$ (see their Figure 2). Using the stellar radial velocity distribution from model 371 (snapshot 680), we found a very similar figure, especially for large velocity values (see top panel of Figure~\ref{fig:DM_density_Kafle}), but an escape velocity of only 321 km $s^{-1}$. The DM mass derived by \citet{Kafle2018} is $M_{\rm 200}$= 7 $\times 10^{11} M_{\odot}$, which is slightly more than twice the value derived from model 371. We suspect that part of the discrepancy is related to an extrapolation of the Navarro-Frenk-White (NFW) profile \citep{Navarro1997}, which requires a very shallow profile at the outskirts (see a discussion about the impact of this extrapolation in \citealt{Jiao2021}). To verify this, we fit model 371's DM-mass radial distribution with a Dehnen profile \citep{Dehnen1993}:

\begin{equation}
\label{eq:dehnen}
    \rho(r)= \frac{M_0\,(3-n)}{4\,\pi\,h^3} \frac{h}{r^n\,(r+h)^{4-n}}.
\end{equation}

We found $M_{0}$=5.93 $\times 10^{11} M_{\odot}$, h=21.64 kpc, and n=1.41 (see red line in the bottom panel of Figure~\ref{fig:DM_density_Kafle}). At a large distance, this density profile follows $\rho$ $\propto$ $r^{-4}$ instead of $r^{-3}$ for the NFW profile, which could partly explain the mass estimate discrepancy. 

The bottom panel of Figure~\ref{fig:DM_density_Kafle} further investigates the origin of the discrepancy between the two mass estimates, and it confirms the fact that even below 15 kpc, \citet{Kafle2018} overestimated the mass (see also their Fig. 6). \cite{Roche2024} summarized the major difficulties in determining mass from escape velocities that are related to (1) the small amount of high-velocity PNe stars (a few tens of stars) used for modeling the tail of the stellar speed distribution and (2) systematics due to strong prior distributions on the fitting parameters, including strong degeneracy between parameters.

The above discussion is necessary for deriving a proper estimate of the DM total mass and its fraction to the dynamical mass. For all methods based on a limited volume such as the HI RC, the substructures surrounding the disk, or the escape velocity from PNe, the DM mass that is calculated has to be within a relatively limited volume, and this follows the definition of what is the missing mass. The latter mass has to be evaluated in a given volume, as does the DM or missing mass fraction. Any extrapolation has to be model dependent, and it should be well described, especially to calculate the DM fraction. It means that within 35 kpc, model 371 has 1.3 and 1.7 $\times 10^{11} M_{\odot}$ for DM and baryonic masses, respectively. Not surprisingly, limiting the volume in which mass is estimated automatically reduces the DM fraction.

A similar high baryonic fraction has been found by \citet{Jiao2023} when estimating the Milky Way mass from the {\it Gaia} DR3 RC. They found a total mass of 2.06 $\times 10^{11} M_{\odot}$ within a radius of 122 kpc, which is made of 1.44 and 0.62 $\times 10^{11} M_{\odot}$ of DM and baryons, respectively. It leads to a baryonic fraction of 30\%, which does not change even when accounting for a smaller radius. Conversely, for M31 our prediction of a baryonic fraction of 32\% within a volume limited by $R_{\rm 200}$ = 137 kpc,\footnote{However, we held to this 32\% baryonic fraction because it directly comes from our modeling that predicts a DM distribution extending up to 137 kpc.} increases to 43\% if it is estimated within 35 kpc. The main reason for this difference is related to the Keplerian decline claimed to be detected by \citet{Jiao2023}, which implies an accurate fit by an Einasto \citep{Einasto1965} profile, while the detection of a Keplerian decline in such a perturbed galaxy such as M31 is excluded.

\section{Conclusions}

In this paper, we have generated a dynamical model of M31 that accounts for a large set of its observed properties, from the bar to the galaxy outskirts. This modeling also succeeds in reproducing its RC detailed properties, namely, the 14 kpc bump and the velocity increase beyond 25 kpc. The former is associated with the axisymmetric disk (see Figures~\ref{RC_Chemin_Corbelli} and ~\ref{RC_asymmetric}), and the latter is caused by returning material from a tidal tail (see Figure~\ref{fig:Angle_s680}) that is out of equilibrium. The above is supported by the fact that for a recent (i.e., $\sim$ 2.6 Gyr old) major merger, gas and stars beyond 30 kpc have no time to experience more than two orbits in the disk or in the halo. 

The DM mass content of our modeling has been calibrated on the M31 RC without affecting its accuracy in reproducing M31's galaxy properties. Our mass estimate supersedes those assuming equilibrium for globular clusters or satellite galaxies at distances well beyond 30-40 kpc, as these probes cannot perform even a single orbit since the merger epoch. In other words, kinematical probes at large radii in such a M31-like galaxy, well beyond 25 kpc, do not trace the underlying mass distribution 
because they are still not at equilibrium with M31's remnant gravitational potential and consequently cannot be relied on to determine its total mass. Given that our modeling accurately reproduces M31's overall properties, including its detailed RC, it also improves estimates based on former analyses of the RC. We find that the M31 RC is consistent with a total baryonic fraction of 32\%, which is far larger than that derived from former studies of external galaxies, including M31. It is also larger than expectations from the cosmic background \citep[baryonic fraction of 15.6\%]{Planck2020}, which casts significant doubt on the existence of a missing baryon problem. 

Finally, we have compared our results to those based on dynamics of substructures in the M31 disk outskirts and to the escape velocity of PNes, and we find that the results are consistent after accounting for their large associated systematics. Further works for consolidating the M31 modeling are very much welcomed. For example, improving the modeling of the central region kinematics could lead to interesting constraints on the DM halo content, which is somewhat linked to the bar properties. A detailed study of kinematical features discovered in the giant stream and in shells \citep{Dey2023} is underway (see \citealt{Tsakonas2024} and a paper in preparation). Other tests are likely to be done by different teams, including about the new Hubble Space Telescope observations of the M31 southern disk \citep{Chen2024}.

\begin{acknowledgements}
We warmly thank the referee for very useful comments. We are grateful for the support of the International Research Program Tianguan, which is an agreement between the CNRS in France, NAOC, IHEP, and the Yunnan Univ. in China. Y.-J.J. acknowledges financial support from the China Scholarship Council (CSC, No.202108070090). LC is grateful to the financial support from Agencia Nacional de Investigacion y Desarrollo of Chile, through the program Fondecyt Regular 1210992. HFW is supported in this work by the Department of Physics and Astronomy of Padova University through the 2022 ARPE grant: {\it Rediscovering our Galaxy with machines.}
      \end{acknowledgements}

%
\bibliographystyle{aa} 
\bibliography{M31_RC.bib} 
%
%
%

\begin{appendix} 
\end{appendix}
\end{document}